\pdfoutput=1
\documentclass[reprint,amsfonts, amssymb, amsmath,  showkeys, nofootinbib,pra, superscriptaddress]{revtex4-2}
\usepackage{float}
\makeatletter
\let\newfloat\newfloat@ltx
\makeatother
\usepackage[english]{babel}
\usepackage[utf8]{inputenc}
\usepackage{graphics}
\usepackage{selinput}

\usepackage[colorinlistoftodos, color=green!40, prependcaption]{todonotes}
\usepackage{braket}
\usepackage{amsthm}
\usepackage{mathtools}
\usepackage{physics}
\usepackage{xcolor}
\usepackage{graphicx}
\usepackage[left=16mm,right=16mm,top=35mm,columnsep=15pt]{geometry} 
\usepackage{adjustbox}
\usepackage{placeins}
\usepackage[T1]{fontenc}
\usepackage{lipsum}
\usepackage{csquotes}

\usepackage{braket}
\usepackage{algorithm}
\usepackage[noend]{algpseudocode}
\def\Cbb{\mathbb{C}}
\def\P{\mathbb{P}}
\def\V{\mathbb{V}}

\def\HC{\mathcal{H}}
\def\KC{\mathcal{K}}

\def\kp{\ket{\psi}}
\def\kpt{\ket{\psi(t)}}
\def\kptN{\ket{\psi_N(t)}}

\def\ad{^{\dagger}}

\def\a{\alpha}
\def\w{\omega}

\usepackage[makeroom]{cancel}
\usepackage[toc,page]{appendix}
\usepackage[pdftex, pdftitle={Article}, pdfauthor={Author}]{hyperref} 

\newcommand{\uba}{Departamento de F\'isica ``J. J. Giambiagi'' and IFIBA, FCEyN, Universidad de Buenos Aires, 1428 Buenos Aires, Argentina}

\newcommand{\losalamos}{Theoretical Division, Los Alamos National Laboratory, Los Alamos, New Mexico 87545, USA}

\begin{document}
\title{Loschmidt echo approach to Krylov-subspace approximation error estimation}

\author{Julian Ruffinelli}
\affiliation{\uba}

\author{Emiliano Fortes}
\affiliation{\uba}


\author{Mart\'{i}n Larocca}
\affiliation{\losalamos}
\affiliation{\uba}

\author{Diego A. Wisniacki}
\affiliation{\uba}

\date{\today} 

\begin{abstract}
The Krylov subspace method is a standard approach to approximate quantum evolution, allowing to treat systems with large Hilbert spaces. 
Although its application is general, and suitable for many-body systems, estimation of the committed error is involved.
This makes it difficult to automate its use. 
In this paper, we solve this problem by realizing that such error
can be regarded as a Loschmidt echo in a tight-binding Hamiltonian. We
show that the different time-regimes of the approximation can be understood using simple physical ideas. More importantly, we obtain computationally cheap error bounds that describe with high precision the actual error
in the approximation.

\end{abstract}


\maketitle

\section{Introduction} \label{sec:introduction}
    
Transmitting and processing information in quantum devices has been established in recent years \cite{bruss2019quantum}. Laboratories around the world are in the race to develop increasingly accurate quantum devices.
To carry out this successfully, it is necessary to test their operation on classical devices.
For this reason, it is important to have efficient classical algorithms to perform quantum simulation \cite{PhysRevLett.88.097904, terhal2002classical}.
    
Several approaches for the efficient computation of quantum time-evolution have been proposed in the literature 
\cite{kluk1986comparison,SCHOLLWOCK201196,PhysRevA.78.012321,daley2004time,PhysRevLett.93.040502}. The cost of the simulation usually depends on specifics of the system, e.g. the initial state, or on the information that
we want to know about the dynamics.
For example, the cost of the simulation can be greatly reduced if the amount of entanglement 
developed by the system remains bounded \cite{daley2004time,PhysRevLett.93.040502}. 
Less restrictive are the well-known Krylov-subspace methods, constructed to provide approximations to the action of the exponential of a matrix on a vector. In the context of quantum simulation, the mechanics of the approximation is the following: an initial state in a (possibly very) large Hilbert space is first mapped to an effective subspace, the Krylov subspace, that captures the most relevant features of the dynamics. Within this low-dimensional subspace, time evolution is (cheaply) computed. Finally, the evolved state is mapped back to the large Hilbert space. Besides quantum simulation, the method has other important applications like solving systems
of ordinary differential equations, large-scale linear systems and more \cite{saad2003iterative,gazzola2020krylov}

The core challenge in Krylov-subspace methods is to keep the error limited, in order to achieve precise evolution. For this reason, it is desirable to be able to predict the time regime in which the error will remain less than a given predetermined tolerance. 
This problem
has been approached in several ways in the literature \cite{park1986unitary,saad1992analysis,stewart1996error,hochbruck1997krylov,expokit,moler2003nineteen,Jawecki:2020cc}, and the provided bounds
generally overestimate the error (significantly). In the seminal paper \cite{park1986unitary}, Park and Light use the fact that the dynamics
in the reduced subspace is that of an effective 1d lattice with a tridiagonal Hamiltonian. An initial state localized at one end starts spreading and the error in the approximation is approximated by the population in the other end of the chain. Later, Saad \cite{saad1992analysis} derived computable estimates of the error using an expansion in the Krylov subspace exploiting the
Lanczos algorithm. Other error bounds include involved computations
making it difficult to use in an operational way \cite{hochbruck1997krylov}.

The goal of this paper is to find tight and computationally inexpensive error bounds for the approximation error in Krylov schemes.
We take advantage of a simple observation: the error can be regarded as a Loschmidt echo in which both the forward and backward evolutions are given by tight-binding Hamiltonians. In a virtual chain, we have an initial state that is localized at one end. The error is related to an echo between evolutions in a $D$ site chain and a trimmed $N<<D$ chain, where $N$ is the dimension of the truncated Krylov subspace used for the approximation.
This analogy allows us to describe the time-regimes of the error using Loschmidt echo theory. In particular, we show that the error remains negligible up to some time at which it starts building up exponentially. This time is related to the effective traveling wave-packet's tail hitting the end of the virtual chain \cite{park1986unitary}. The core of our proposal is that the error in this regime can be captured remarkably well by replacing the full-size backwards evolution with one of a chain with only an single extra site. This provides an accurate and cheap bound for the error.

Moreover, we show that one can analytically solve for the
bound in the case in which the tight-binding Hamiltonian has 
homogeneous diagonal and off-diagonal elements. We test this solution in a 1-D Ising spin chain with transverse magnetic field. Finally, we give some physical insight
explaining why this simple model works in the general case.

The paper is organized as follows. In Sec. \ref{sec:Krylov}, we introduce the general framework of the Krylov-subspace method for quantum time evolution. Next, in Section. \ref{sec:regimes} we describe the different time-regimes of the error, focusing on the analogy with Loschmidt echo dynamics under tight-binding Hamiltonians.
In section \ref{sec:numres}, we use the connection between the error and the Loschmidt echo of tight-binding Hamiltonians to propose a bound that
describes extremely well the inaccuracy of the approximate evolution in the Krylov subspace.
Finally, in Sec.\ref{sec:conclu} we offer some final remarks. Appendix \ref{sec:ap_lanczos} provides a brief description of Lanczos algorithm and in Appendix \Ref{sec:ap_ising} we describe the system used for the numerical simulations, a 1-D Ising spin chain with transverse magnetic field. 
In Appendix \ref{sec:analytical}, the error bound is analytically solved for
the simple case in which the tight-binding Hamiltonian has homogeneous diagonal and non-diagonal elements.


\section{The Krylov-subspace Method} \label{sec:Krylov}

Let us start by reviewing the so-called Krylov-subspace method for approximating quantum dynamics. Consider a state $\kp$, in a $D$-dimensional Hilbert space $\HC=\Cbb^D$, that evolves under a time-independent Hamiltonian $H$. The $N$-dimensional Krylov subspace associated with $\kp$ and $H$ is given by

\begin{equation}\label{eq:kry_subs}
    \KC_N = \text{span} \{ \kp , H \kp , \ldots, H^{N-1} \kp \}
\end{equation}
Here, without loss of generality, we consider that $H$ and $\kp$ share no symmetries, i.e. such that $\KC_D=\HC$. If they did share some symmetry, time evolution would occur constrained to a symmetry subspace, say $\HC_j \subset \HC$, in which case one can always redefine the problem to belong within that subspace, e.g. $\HC \gets \HC_j$. 

\begin{figure}[htbp]
\centerline{\includegraphics[scale=.3]{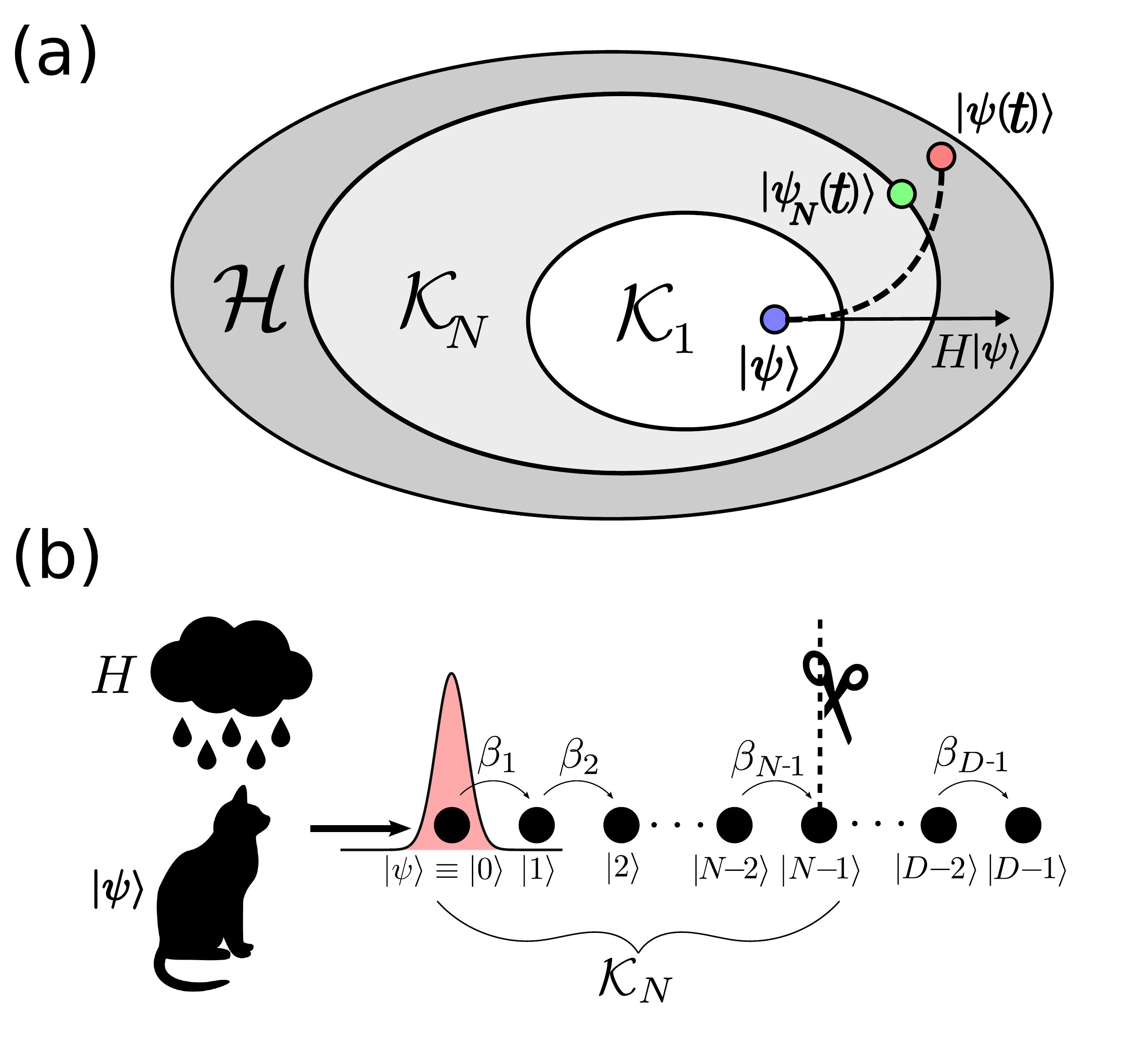}}
\caption{\textbf{Schematic Krylov Approximation:} (a) An initial state $\kp$ (blue circle) evolves under Hamiltonian $H$, drawing some trajectory on Hilbert space $\HC$ (dashed line). At time $t$, the evolved state is $\kpt$ (red circle). The Krylov approach consists in approximating this state with $\kptN$, its projection into the Krylov subspace $\KC_N$ (green circle), defined in Eq.~\eqref{eq:kry_subs}. (b) The dynamics of $\ket{\psi}$ under $H$, from the Lanczos Basis perspective, corresponds to the diffusion of an initial state $\ket{0}$ that is completely localized at the leftmost end of a virtual tight-binding chain. Here, the off-diagonal elements of Lanczos tridiagonal matrix, $\beta_i$, act as hopping amplitudes between neighbouring sites and the diagonal elements $\alpha_i$ as local onsite potentials (not depicted in the image). Using a truncated Lanczos basis can be regarded as "cutting" the chain at site $N$.}
\label{fig:kry}
\end{figure}

The Krylov approach aims at approximating the time-evolved state $\kpt$ with the best element $\kptN \in \KC_N$. To do so, we first have to build an orthonormal basis for $\KC_N$, $B_N=\{ \ket{v_0}\equiv \kp,\ldots, \ket{v_{N-1}}\}$. This is usually done using Lanczos's algorithm, a sort of Gram-Schmidt procedure that harnesses the fact that orthonormalization only needs to be enforced with respect to the last two vectors in the basis (see Appendix \ref{sec:ap_lanczos}). Once we have a basis for $\KC_N$, we can get an approximation for $\kpt$ by projecting it into this basis (see Fig. \ref{fig:kry} (a) for a schematic representation of the method)

\begin{equation}
\begin{split}
    \kpt=e^{-i H t} \kp &\approx \P_N e^{-i H t} \P_N \kp    \\
    & =\V_N\ad e^{-i T_N t} \V_N \kp\\
&\equiv \kptN
\end{split}
\label{krylov}
\end{equation}
Here, $T_N = \V_N H \V_N\ad$ is the Hamiltonian reduced to the subspace, $\V_N$ and

\begin{equation}
    \V_N\ad = \begin{bmatrix} \vdots &  \vdots  && \vdots \\ \ket{v_0}, & \ket{v_1},&  &, \ket{v_{N-1}} \\ \vdots &  \vdots  && \vdots \end{bmatrix} 
\end{equation}
are the reduction-to-the-subspace operators, and $\P_N=\V_N\ad \V_N$ the projector onto it. By definition, $\V_N$ maps any initial state into the first coordinate vector of an effective $N$-dimensional system, $\V_N\kp=(1,0,\cdots,0)^T\equiv \ket{0}_N$. It is especially relevant to notice that the Hamiltonian reduced to a Krylov subspace is tridiagonal

\begin{equation}
\label{eq:tridiag}
T_N =    \begin{pmatrix}\alpha_{1} & \beta_{1} & 0 & \cdots & 0\\
\beta_{1} & \alpha_{2} & \beta_{2} & \cdots & 0\\
0 & \beta_{2} & \alpha_{3} & \cdots & 0\\
\vdots & \vdots & \vdots & \ddots & \vdots\\
0 & 0 & 0 & \cdots & \alpha_{N}
\end{pmatrix}
\end{equation}
and thus, this effective system has the form of a tight-binding model. An initial state localized in one end of an effective chain evolves according to $T_N$ (i.e. with onsite potential $\a_i$ and hopping amplitude $\beta_i $ at the ith site) propagating the excitation and populating the rest of the lattice (see Fig. \ref{fig:kry}(b) for a schematic representation).  Finally, $\V_N\ad$ maps the effective evolved state back to full Hilbert space. The efficiency of the method resides in the fact that the time evolution is solved "cheaply" in the reduced space, i.e. one replaces the exponential of a $D\times D$ Hermitian matrix $H$ with the much more economical exponential of a $N\times N$ symmetric tridiagonal $T_N$. Of course, the assumption is that $N << D$


The challenge in this approximate evolution scheme is to keep the error bounded by a given tolerance. This has been studied in different ways for more than three decades \cite{park1986unitary,saad1992analysis,stewart1996error,hochbruck1997krylov,expokit,moler2003nineteen,Jawecki:2020cc}. In the next Section, we show that the error as a function of time has regimes that can be well understood using physical ideas based on Loschmidt echo theory and diffusion in a tight-binding model \cite{Goussev:2012}.


\section{time regimes of the error} \label{sec:regimes}




Let us review the time regimes of the error in the Krylov-subspace method. This error is given by the instantaneous infidelity between exact and approximate evolved states


\begin{equation}\label{eq:error}
   \epsilon_N(t) = 1-|\braket{\psi_{N}\left(t\right)}{\psi \left(t\right)}|^2
\end{equation}
Any actual implementation of the approximation method has to keep track of this error. Yet, of course, it's exact computation is out of question since it involves solving the problem one is trying to approximate, $\ket{\psi(t)}$.

A closer inspection of Eq.~\eqref{eq:error} allows for an interesting interpretation. Rewriting the overlap as



\small
\begin{equation}
\begin{aligned}
   |\braket{\psi_{N}\left(t\right)}{\psi \left(t\right)}|^2
    &=\big|\bra{\psi} \V_N^{\dagger} e^{i T_N t} \V_N e^{-iHt} \ket{\psi}\big|^2\\
    &=\big|\bra{\psi}  \V_N^{\dagger} e^{i T_N t} \V_N \V_D^{\dagger} e^{-i T_D t} \V_D| \ket{\psi}\big|^2\\
    &=\big|\bra{\psi} \V_D^{\dagger} e^{i \tilde{T}_N t} \V_D \V_D\ad e^{-i T_D t} \V_D \ket{\psi}\big|^2\\&=
    \big|\bra{0}  e^{i \tilde{T}_N t} e^{-i T_D t} \ket{0} \big|^2
\end{aligned}
\label{echo-binding}
\end{equation}
\normalsize 
where $\tilde{T}_N = \V_D \P_N H \P_N \V_D\ad$ has the form 



\begin{equation}
\tilde{T}_N=\left(\begin{array}{l|l}
T_N & 0 \\
\hline 0 & 0
\end{array}\right)
\end{equation}
one can realize that $1-\epsilon_N(t)$ has the form of a Loschmidt echo \cite{Goussev:2012} in which the backwards and forward evolutions are given by tight-binding Hamiltonians. We start with $\ket{0}\equiv \V_D \kp$, a completely localized state at one end of the virtual chain. This state evolves subject to $T_D$ for some time t, then evolves backwards subject to $\tilde{T}_N$ (a perturbed $T_D$ where the effective onsite potentials and hoppings of sites $N+1,\ldots,D$ are turned off) and is finally overlapped.

The Loschmidt echo has been widely studied as a measure of the revival occurring after a forward and backward time evolutions generated by two slightly different Hamiltonians \cite{gorin2006dynamics,jacquod2009decoherence,Goussev:2012} . As far as we know, the case of tight-binding Hamiltonians has not been explicitly considered in the literature so far. We note that one of the evolutions is done with a chain of length $D$ and the other evolution corresponds to the case in which the chain is cut at site $N$ (the hoppings and onsite potentials at the second part of the chain are set to zero, i.e. $\alpha_i=0$ and $\beta_i=0$  for $i=N+1,...D$). 

\begin{figure}[htbp]
\centerline{\includegraphics[scale=.37]{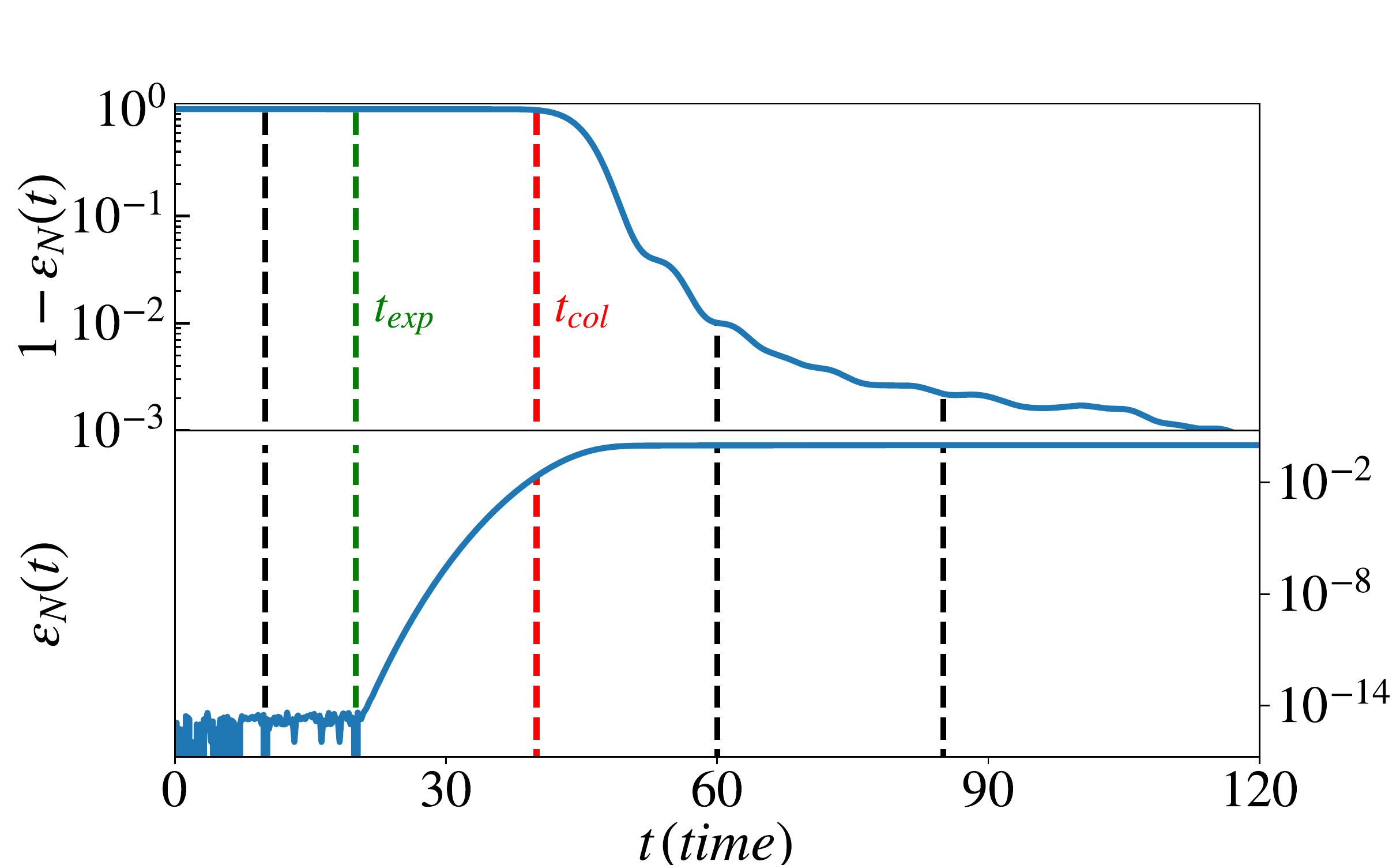}}
\caption{\textbf{Time regimes of the echo.} Loschmidt echo 
$|\braket{\psi_{N}\left(t\right)}{\psi \left(t\right)}|^2=|\braket{\psi_{N}\left(t\right)}{\psi_D \left(t\right)}|^2$
(top panel) and error $\epsilon_N(t)$ (bottom panel) for an Ising spin chain with transverse magnetic field.
We use $N=30$ and $D=2^n=1024$.
We have marked with dashed vertical lines the times that correspond to the snapshots shown in Fig. \ref{fig:snapshots}.
We also highlight the relevant times $t_{exp}$ and $t_{col}$.  The initial state $\ket{\psi}$ is random state.}
\label{fig:echo}
\end{figure}

In order to gain insight on the time regimes of the approximation,
we show in Fig. \ref{fig:echo} the Loschmidt echo $|\bra{\psi_{N}\left(t\right)}\ket{\psi \left(t\right)}|^2$ (top panel) and the error $\epsilon_N(t)$ (bottom panel) for an Ising spin chain with $10$ sites and a transverse magnetic field (see Sec. \ref{sec:numres} for more details). 
We use a Krylov-subspace of $N=30$ sites
and a random initial state $\ket{\psi}$. We can clearly
see that the Loschmidt echo has two very different time regimes. Until $t \approx t_{col}$ the echo remains roughly one and the approximate evolution faithfully captures the exact one.  After this first "faithful" regime, an abrupt decrease is observed and from there on it decays in a monotonous way.  

In this first time-regime $t<t_{col}$ where the echo practically does not change, the error has two relevant regimes. First, until some time $t<t_{exp}$, the error is essentially zero. Then, at $t=t_{exp}$ the error suddenly starts to build-up exponentially. This is related to the tail of the wave-packet starting to impact on the end of the chain. The interval $ t_{exp} \le t \le t_{col}$ is precisely the region that we have to be able to correctly approximate in order to have a good error bound. We note that the noisy plateau of $\varepsilon_N(t)$ for $t<t_{exp}$ is due to round-off errors in the floating-point arithmetic used in the computations.

\begin{figure}[htbp]
\centerline{\includegraphics[scale=.25]{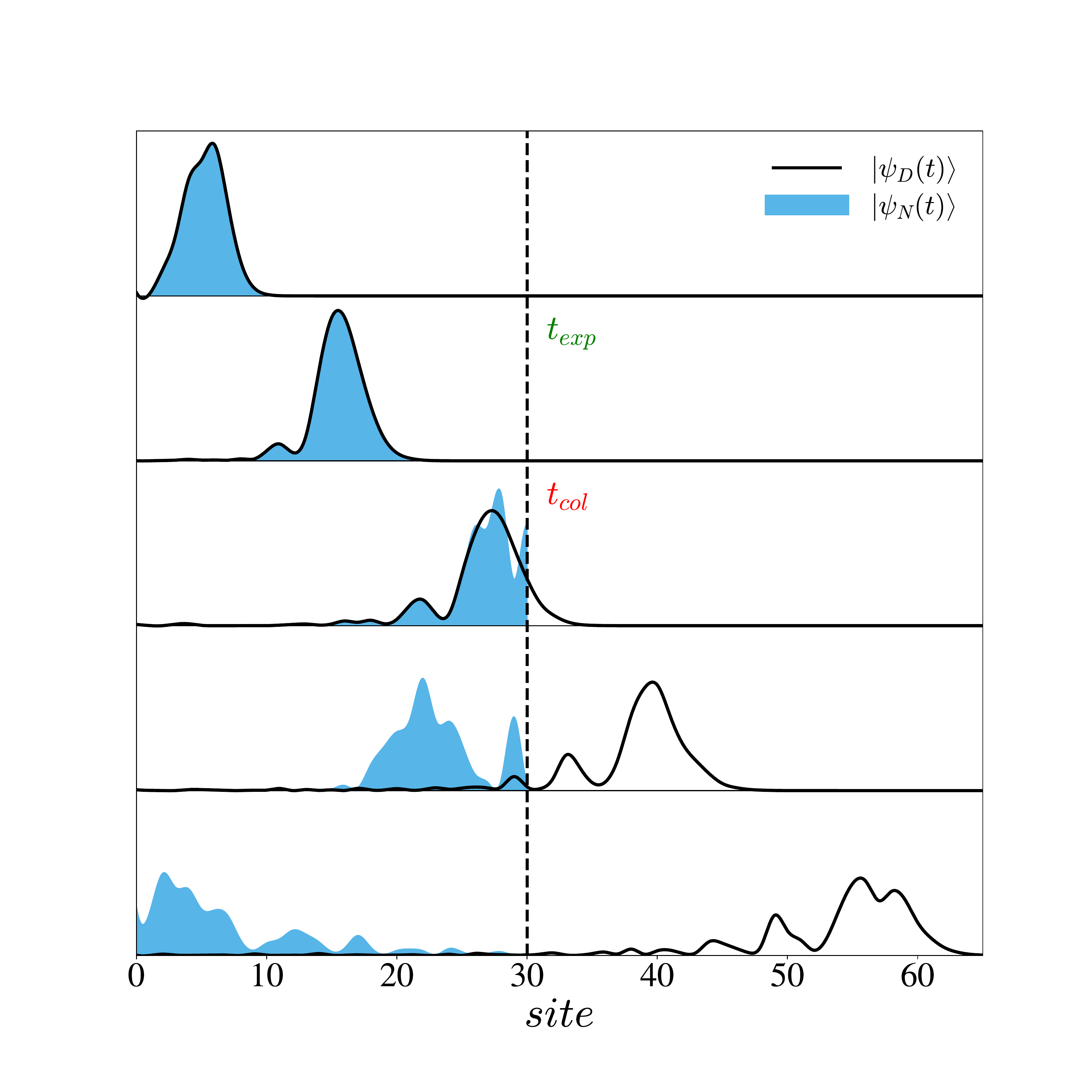}}
\caption{\textbf{Time evolution of exact and approximate states in the Lanczos basis.} We draw $|\braket{\psi_{D}(t)}{v_i}|^2$ (black line) and $|\braket{\psi_{N}(t)}{v_i}|^2$ (solid blue) at times $t=10, 25, 42, 60$ and $80$ (top to bottom). Remark: the representation in the figure takes a cubic interpolation between each site to smooth out the discrete sites effect for an easier visualization. }
\label{fig:snapshots}
\end{figure}

In order to understand the time regimes of Fig. \ref{fig:echo}, we plot in Fig. \ref{fig:snapshots} the square of the amplitudes in the Lanczos basis for both the exact and approximate evolved states of Eq.
(\ref{echo-binding}), i.e. 
$|\Braket{v_i|\psi_{N}(t)}|^2$ and $|\Braket{v_i|\psi_{D}(t)}|^2$ for $i=1,...N$ and  $i=1,...D$, respectively. We provide snapshots of this virtual travelling wave-packets at times $t=10, 25, 45$ and $70$.  As mentioned before, we start with localized states at one end of the fictitious tight-binding chain. In the first panel of Fig. \ref{fig:snapshots}, corresponding to time $t=10$, both wave-packets are travelling to the right and are essentially equal. However, at $t=t_{exp} \approx 25$, the exponential tail reaches the site $N = 30$, and the error starts to build-up rapidly. This process continues until $t=t_{col}\approx42$, where one of the packets bounces with the end of its chain and starts returning to its original position. This difference in the behavior of the wave-functions is reflected in an abrupt decay of the echo (see Fig. \ref{fig:echo}). At $t=60$ and $t=80$, the wave packets continuously grow apart and become more and more orthogonal.

.

\section{From Loschmidt echoes to error bounds} 
\label{sec:numres}

In the previous section, we have shown that the error in the Krylov
method for quantum evolution can be seen as a Loschmidt echo. 
Let us now show how this description can help us derive tight and
computationally cheap bounds for the error, providing a fundamental 
tool for any practical implementation of the approximation method.

\begin{figure}[htp]
\includegraphics[scale=.38]{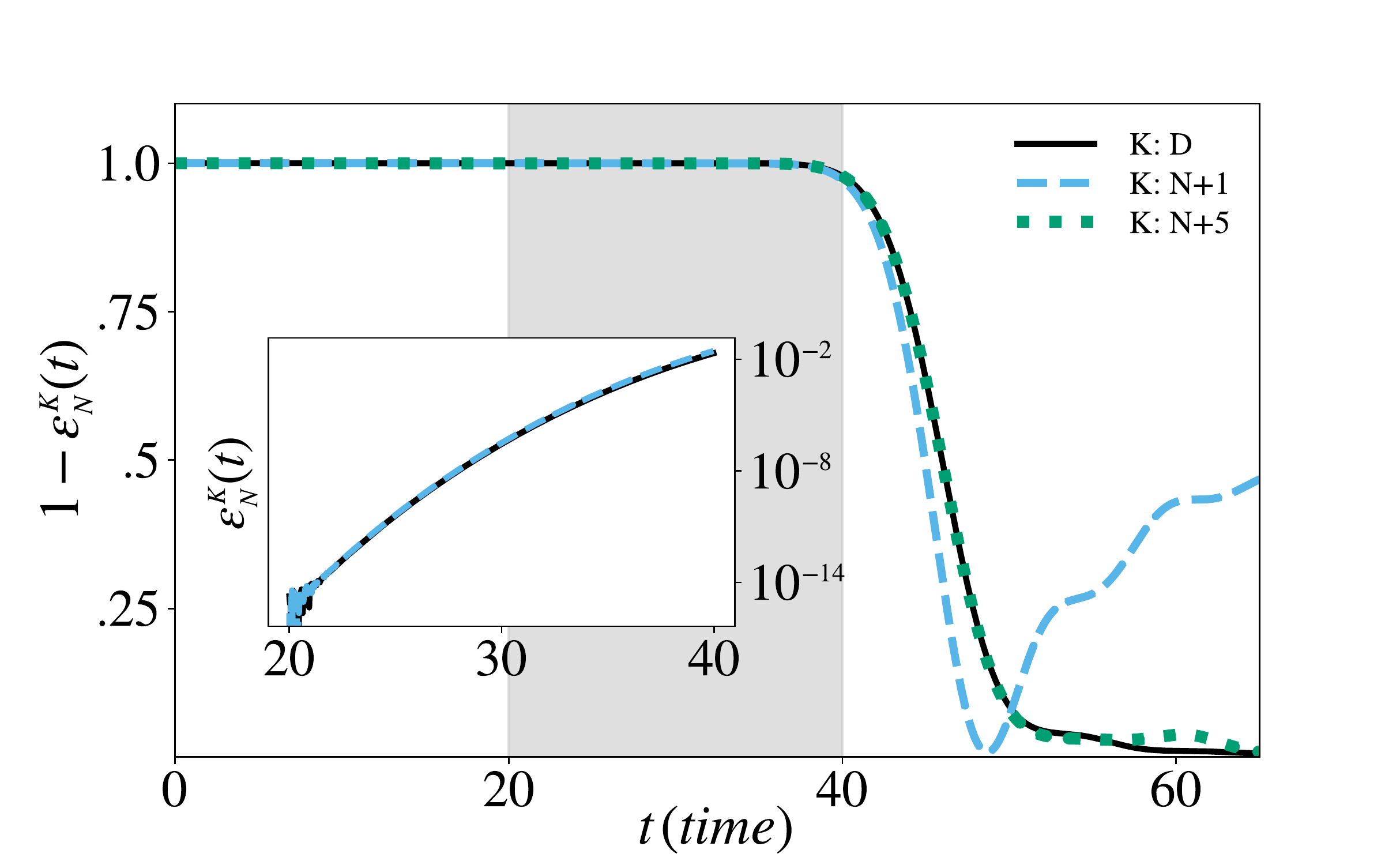}
\caption{Loschmidt echo $|\braket{\psi_{N}\left(t\right)}{\psi_{K}\left(t\right)}|^2$ with $K=N+1$ (light blue dashed line), $K=N+5$ (green dotted line) and $D$ (black solid line). Here, we use $D=2^{10}$-dimensional Ising spin chain with transverse magnetic field. \textbf{Inset}: The error  $\varepsilon_N^{K}(t)$ in the shaded region of the main plot.  }
\label{fig:echo2}
\end{figure}

As we have previously remarked, the time-regime of the error that is relevant for a practical implementation of the approximation method is the one between $t_{exp}$ and $t_{col}$.
In this region, the travelling packet has its center between sites $1$ and $N$, and only a small, exponentially suppressed population tail surpasses the site $N$. With this in mind, we can ask ourselves: is it really necessary to consider the entire chain to describe the behavior of the error? Given that in the $[N+1,D]$ region we have exponentially suppressed populations, isn't it possible to capture the essential features of the error by considering instead an echo where we replace the full chain with one with $K=N+i$ sites, i.e.
where $i$ is a small number of extra sites?  To answer this question, in Fig. \ref{fig:echo2}, we compare the echo  $|\bra{\psi_{N}\left(t\right)}\ket{\psi\left(t\right)}|^2$ with  $|\bra{\psi_{N}\left(t\right)}\ket{\psi_{K}\left(t\right)}|^2$ using $K=N+1$ and $K=N+5$. Here, we can see that both cases accurately capture the important region between $t_{exp}$ and $t_{col}$
(shaded region of Fig. \ref{fig:echo2}). In the inset of Fig. \ref{fig:echo2} we plot the error $\varepsilon_N^{K}(t)=1-|\bra{\psi_{N}\left(t\right)}\ket{\psi_{K}\left(t\right)}|^2$ in the shaded region to highlight this last conclusion.
This remarkable fact, i.e. that only a singe extra site is enough to capture the behaviour of the error in the relevant region, will be the main building-block of our error bound proposal. Let us note that, although here we have used a spin chain model, the systematic appears to be the same for other fundamentally different systems \cite{new}. We have tested in random Hamiltonians that belong to the Gaussian orthogonal ensamble (GOE) and Gaussian unitary ensamble (GUE)  \cite{edelman2005random}. 

Let us suppose that we have computed the Krylov subspace $\KC_N$ and want to estimate the error in this case. As we have argued in the previous paragraph, one can effectively approximate the error with $|\bra{\psi_{N}\left(t\right)}\ket{\psi_{N+1}\left(t\right)}|^2$. In order to do so, one would have to perform an extra iteration of the Lanczos algorithm, i.e. to compute this extra site approximation $\ket{\psi_{N+1}(t)}$. 
Alternatively, let us see if it is possible to approximate the new site in the tight-binding chain without having to do such extra iteration.
One simple yet effective way of estimating the coefficients of this new site is to average over the
previous sites.
That is,
\begin{equation}
    \alpha_{N+1}\approx \bar{\a}\equiv \frac{1}{N}\sum_1^{N} \alpha_i,\quad\quad
    \beta_{N+1}\approx\bar{\beta}\equiv \frac{1}{N}\sum_1^{N} \beta_i
\label{app}
\end{equation}

\begin{figure}[htbp]
\centerline{\includegraphics[scale=.38]{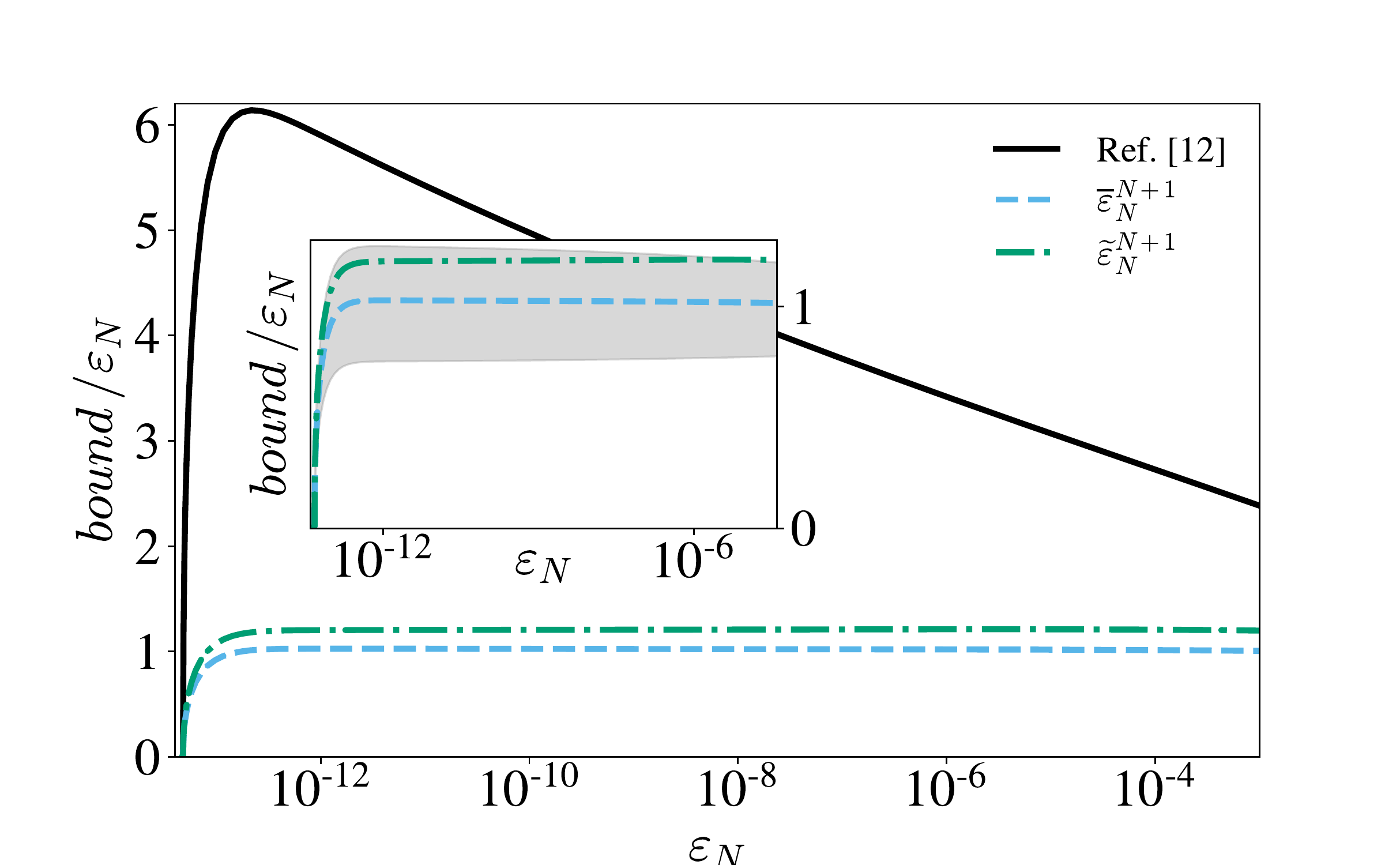}}
\caption{Ratio of the bounds $\bar{\varepsilon}_N^{N+1}$
(dashed line), $\tilde{\varepsilon}_N^{N+1}$ (dash-doted line) and the a posteriori bound of
Ref. \cite{saad1992analysis} (solid line) with the the actual error $\varepsilon_N$ vs. $\varepsilon_N$. See text for more details. }
\label{fig:bounds}
\end{figure}

Now, we have all the elements needed to test our bound and to compare it with the established bounds in the literature, e.g. of Ref. \cite{saad1992analysis}.
In Fig. \ref{fig:bounds}, we show the ratio between the error bounds $\bar{\varepsilon}_N^{N+1}$ and the actual error $\varepsilon_N$ of Eq.~\eqref{eq:error}.
We put a bar on the bound $\bar{\varepsilon}_N^{N+1}$ to denote that we use the averaged estimation of Eq.~\eqref{app} for the coefficients of site $N+1$.
In the inset of \ref{fig:bounds},
we shade the region of the bound $\varepsilon_N^{N+1}$ in which the elements $\alpha_{N+1}$ and $\beta_{N+1}$ are the maximum o the minimum of $\alpha_{i}$ and $\beta_{i}$ with
$i=1,...N$.
We also show the ratio of the bound of Ref. \cite{saad1992analysis} with the actual error.
It can be seen that in both of the bounds proposed the overestimation
remains constant throughout the evolution, and quite lower than Ref. \cite{saad1992analysis}.

Interestingly, the echo $|\bra{\psi_{N}\left(t\right)}\ket{\psi_{N^\prime}\left(t\right)}|^2$  can be solved analytically in the
particular case of homogeneous coefficients,
$\alpha_i = \alpha$ and $\beta_i = \beta$ $\forall i$, which corresponds to the Toeplitz tridiagonal matrix \cite{toeplitz_matrix} (see appendix \ref{sec:analytical}).
Using this analytic expression, we compute a new bound $\tilde{\varepsilon}_N^{N+1}$ (with a tilde) where we use Eq.~\eqref{eq:topelitz_echo} with $\a=\bar{\a}$ and $\beta=\bar{\beta}$ in Eq.~\eqref{app}. We show
in Fig.
\ref{fig:bounds} that this approximation works also very well.

\section{Conclusions} \label{sec:conclu}


In this work, we have established a connection between the behaviour of the error of Krylov-subspace approximations for quantum time evolution and a Loschmidt echo between effective wave-packets travelling in virtual chains with nearest-neighbour hopping amplitudes. One of the chains has $D$ sites and the other one $N<<D$. The packages start-off at the leftmost end of the chain, and during some time their profile is identical. Then at $t\approx t_{exp}$ one of the packets tail starts colliding with the end of the chain, and error commences to build-up exponentially. At a later time $t\approx t_{col}$ the center of this packet arrives at the end of the chain and bounces back, whereas the other packet still travels undisturbed. Here, the error reaches significant values and the echo departs from unity. Hereafter, the packages travel in opposite directions and they become evermore 
orthogonal.

In practice, any approximation method must be accompanied with an efficient and accurate error estimator. Now, error estimation for Krylov-subspace method has been an elusive subject for more than 30 years \cite{park1986unitary,saad1992analysis,stewart1996error,hochbruck1997krylov,expokit,moler2003nineteen,Jawecki:2020cc}. In this context, and besides providing a nice physical-insight on the mechanics of the error, the Loschmidt echo picture offers an elegant and simple solution to the error tracking problem. Remarkably, we show that one can capture with extreme precision the behaviour of the error in the relevant region, almost without having to make any extra computations.

Typical implementations of Krylov-subspace methods usually involve a time-stepping schedule \cite{expokit}. The reason for this is that Lanczos's Algorithm suffers from instabilities when the basis is too big. Thus, the common workaround is to approximate the evolution using an iterative approach: the actual trajectory in Hilbert space is efficiently followed using a sequence of patches. That is, we build a Krylov-subspace, evolve for a small time, map back and start over. In this framework, our error bounds provide cheap and accurate way of computing optimal time intervals for the time-stepping schedule.





\section*{Acknowledgements} \label{sec:acknowledgements}
The work was partially supported by CONICET (PIP 112201 50100493CO), UBACyT (20020130100406BA), ANPCyT (PICT-2016-1056), and Unitary Fund. ML was supported by the U.S. Department of Energy (DOE), Office of Science, Office of Advanced Scientific Computing Research, under the Accelerated Research in Quantum Computing (ARQC) program as well as under the Quantum Computing Application Teams program. 

\appendix
\section{Lanzcos method}\label{sec:ap_lanczos}

The Lanczos method (see Algorithm \ref{lanczsos}) is a well-know strategy for the construction of $B_N=\{\ket{v_0},\dots,\ket{v_{N-1}}\}$,
an orthonormal basis spanning the Krylov subspace $\KC_N$. One of the most appealing features of this approach is that,
unlike e.g. a Gram-Schmidt procedure where orthonormalization at each step is with respect to the whole current basis, the new candidate vector $\ket{x_j}$ only needs to be orthonormalized with respect to the previous two basis vectors $\ket{v_{j-1}}$ and $\ket{v_{j-2}}$.
The reason for this is that the Hamiltonian,
by construction, is tridiagonal in the Lanczos Basis (see Eq. \eqref{eq:tridiag}).

\begin{algorithm}
    \label{lanczsos}
    \begin{algorithmic}[1]
        \State $\ket{v_0}=\kp$ (assume normalized)
        \State $\ket{x_1}= H \kp$
        \State $\a_1= \braket{x_1}{v_0}$ (the component of $\ket{x_1}$ in $\ket{v_0}$)
        \State $\ket{w_1}=\ket{x_1}-\a_1\ket{v_0}$
        \For {$j = 1,2,\ldots$}
            \State $\beta_j=\sqrt{\braket{\w_j}{\w_j}}$
            \If{$\beta_j>0$}
                \State $\ket{v_j}\gets\frac{1}{\beta_j}\ket{\w_j}$.
                \Else 
                \State break
            \EndIf
            \State $\ket{x_{j+1}}= H\ket{v_j}$
            \State $\a_{j+1}= \braket{x_{j+1}}{v_j}$
            \State $\ket{\w_{j+1}}=\ket{x_{j+1}}-\a_{j+1}\ket{v_j}-\beta_j\ket{v_{j-1}}$
        \EndFor
    \end{algorithmic}
\caption{Lanczos Algorithm. Receives state $\kp$ and Hamiltonian $H$ and returns a set of $N$ orthonormal vectors $\{\ket{v_i}\}$ spanning the Krylov subspace $\KC_N$.}
\end{algorithm}

\section{Ising spin chain in a transverse magnetic field}\label{sec:ap_ising}
Let us describe the system used in the numerical simulations. Consider a 1D Ising spin chain with transverse magnetic field and open boundary conditions, described by, 

\begin{equation}\label{hamilt}
H=\sum_{k=1}^{N}\left(h_x \hat{\sigma}_{k}^{x}+h_z\hat{\sigma}_{k}^{z} \right) -  J \sum_{k=1}^{N-1} \hat{\sigma}_{k}^{z}\hat{\sigma}_{k+1}^{z},
\end{equation}

where $N$ is the total number of spin-$1/2$ sites of the chain,
$\hat{\sigma}_{k}^{j}$ to the Pauli operator at site $k=\lbrace1,2,...,N\rbrace$
with direction $j=\lbrace x,y,z \rbrace$ and $J$ represents the interaction strength within the site $k$ and $k+1$.
The parameters $h_x$ and $h_z$ are,respectively,
the strength of the magnetic field in the (transverse) $x$ direction,
and in the (parallel) $z$ direction.
We set $\hbar=1$, such that energies are measured in units of the interaction strength $J$, and times in units of $J^{-1}$. \\
\\

\section{ Analytical solution of the error: special case of homogeneous hopping} \label{sec:analytical}

We solve here a simplified model for the evolution on the Krylov subspace $\KC_N$. Let us assume that after mapping $\kp$ and $H$ to $\ket{0}_N$ and $T_N$, we find a homogeneous tridiagonal matrix,

\begin{equation}\label{eq:topelitz_hamiltonian}
    T_N=\a \sum_{n=1}^{N} \ket{n} \bra{n} + \beta \sum_{n=1}^{N-1} \ket{n} \bra{n+1}+\text{h.c.}.
\end{equation}
Here, $\ket{n}\equiv \ket{n}_N$ (here and hereafter we drop de subscript) denotes the localized "site" states of the $N$-dimensional tight-binding chain associated with the dynamical system ($\psi$,$H$). The Hamiltoinan in Eq. ~\eqref{eq:topelitz_hamiltonian} corresponds to the so-called Toeplitz tridiagonal matrix \cite{toeplitz_matrix}, and has well documented analytical expressions for its eigenstates and eigenenergies,

\begin{equation}
    \label{eq:topelitz_eigenvectors}
\left\langle n | E_{k}\right\rangle =\sqrt{\frac{2}{N+1}}\sin\left(\frac{nk\pi}{N+1}\right),
\end{equation}

and 

\begin{equation}
        \label{eq:topelitz_eigenvalues}
E_{k}=\alpha+2 \beta \cos \left(\frac{n k \pi}{N+1}\right).
\end{equation}

The time evolution of an arbitrary initial state $\ket{\psi(t=0)}=\sum_{n=1}^N c_n \ket{n}$ is given by,

\begin{equation}
    \label{eq:topelitz_evolution}
    \begin{aligned}
        \ket{\psi(t)} & =\sum_{n,n'}^{N}c_{n} S_{n,n'}^{N}(t)\ket{n'},
    \end{aligned}
\end{equation}
where the transition matrix $S_{n,n^{\prime}}^N(t)$ is defined as, 

\begin{equation}
    \label{eq:topelitz_transition_matrix}
    S_{n,n^{\prime}}^N(t) = \sqrt{\dfrac{2}{N+1}} \sum_{k=1}^{N}
    \sin\left(\frac{nk\pi}{N+1}\right)
    \sin\left(\frac{n'k\pi}{N+1}\right)
    e^{itE_{k}}.
\end{equation}

Finally, the amplitude of the echo of two time-evolutions with Toeplitz matrices of lengths $N$ and $N'$ yields,


\begin{equation}
    \label{eq:topelitz_echo}
 \Bra{0}
    e^{-i t T_{N}^{\prime}} e^{i t T_{N}}
    \Ket{0} =1-
    \sum_{n=1}^{N^{\prime}} 
    S_{1 n}^{N}(t) S_{n,1}^{N^{\prime}}(-t).
\end{equation}

 It is clear from  Eq.~\eqref{eq:topelitz_eigenvalues}, that the parameter $\alpha$ will not affect the value of the echo and $\beta$ acts as a rescaling of time. Thus, one can limit itself to study the behavior of the chain with parameters $\alpha=0$ and $\beta=1$, and then rescale time by $\beta t$.




\bibliography{main.bib}


\end{document}